# AMBRE: A COMPACT INSTRUMENT TO MEASURE THERMAL IONS, ELECTRONS AND ELECTROSTATIC CHARGING ONBOARD SPACECRAFT

B. Lavraud[1], A. Cara[1,2], D. Payan[2], Y. Ballot[3], J.-A. Sauvaud[1], R. Mathon[1], T. Camus[1], O. Chassela[1], H.-C. Seran[1], H. Tap[4], O. Bernal[4], M. Berthomier[5], P. Devoto[1], A. Fedorov[1], J. Rouzaud[1], J. Rubiella-Romeo[6], J.-D Techer[5], D. Zély[6], S. Galinier[6], and D. Bruno[7]

[1] IRAP, CNRS, CNES, Université de Toulouse, France
[2] CNES, Toulouse, France
[3] EREMS, Flourens, France
[4] LAAS-CNRS, Université de Toulouse, CNRS, INPT, Toulouse, France
[5] LPP, Paris, France,
[6] Mécano ID, Toulouse, France,
[7] COMAT, Flourens, France

## ABSTRACT

The Active Monitor Box of Electrostatic Risks (AMBER) is a double-head thermal electron and ion electrostatic analyzer (energy range 0 – 30 keV) that was launched onboard the Jason-3 spacecraft in 2016. The next generation AMBER instrument, for which a first prototype was developed and then calibrated at the end of 2017, constitutes a significant evolution that is based on a single head to measure both species alternatively. The instrument developments focused on several new sub-systems (front-end electronics, high-voltage electronics, mechanical design) that permit to reduce instrument resources down to ~ 1 kg and 1.5 W. AMBER is designed as a generic radiation monitor with a twofold purpose: (1) measure magnetospheric thermal ion and electron populations in the range 0-35 keV, with significant scientific potential (e.g., plasmasphere, ring current, plasma sheet), and (2) monitor spacecraft electrostatic charging and the plasma populations responsible for it, for electromagnetic cleanliness and operational purposes.

## 1. INTRODUCTION

Plasmas constitute the vast majority of the mass of the visible universe. Knowledge of their properties and of the key plasma processes at work is fundamental to understanding their dynamics. Near-Earth plasmas have long been used as a natural laboratory for the study of astrophysical plasmas. In addition, knowledge of the plasma environment is critical to the understanding of its interaction with spacecraft, with both scientific (plasma-material interaction science) and operational purposes (radiation effects on spacecraft systems). Concerning operational aspects, both high and low-energy particle radiations have effects on spacecraft. While high energy particles lead to damage on spacecraft electronics (through overall dose effects, single event upsets, etc.), low energy particles do not penetrate deep into materials but they do alter spacecraft surface properties through their absorption and the settling of a surface electrostatic potential.

The presence and intensity of the electrostatic charging of spacecraft surfaces is of course critical for electromagnetic cleanliness (as it affects the electric properties of the spacecraft) but also to overall spacecraft health as this charging is often differential (not the same on all part of the spacecraft), thereby potentially leading to discharges with detrimental impacts on spacecraft systems.

Better understanding and mitigating spacecraft charging effects require the measure of the charging onboard spacecraft. This may be done with dedicated instruments as presented here. Such instruments, in addition, provide measurements of the ion and electron populations at the origin of this charging. Indeed, spacecraft electrostatic charging results from a subtle equilibrium between currents generated by various sources: solar UV flux (inducing a photoelectron currents), secondary electrons (from thermal electron interaction with the spacecraft surfaces), and ambient ion and electron populations (producing currents flowing towards/away from the spacecraft surface).

Appropriate thermal plasma measurements have been performed on many spacecraft, but most were dedicated to science, for which electromagnetic cleanliness is of particular importance. On other spacecraft, e.g. commercial or strategic satellites, such measurements are rarely performed despite their obvious importance for measuring and mitigating the impact of spacecraft charging. For instance, a number of DOE (Department Of Energy) satellites in geosynchronous orbit possess instruments measuring ions and electrons up to 40 keV on-board each spacecraft: the Los Alamos National Laboratory (LANL) Magnetospheric Plasma Analyzer (MPA) [1,2]. The geosynchronous orbit is of course a very important one, given its considerable use for commercial (e.g. telecom) and defence needs. The strategic nature of the LANL MPA measurements led this dataset to be removed from public release since 2008. Another rare dataset, still accessible, is that from the DMSP Sun-synchronous satellites, in a polar orbit at roughly 835 km, which also provides a considerable set of ion and electron measurements up to 30 keV [3].

Despite their obvious interest, at present and to our knowledge, such measurements are not systematically performed on any major suite of non-scientific spacecraft in Europe.

Here we present the recent and ongoing developments of the AMBRE instrument, which aim at providing a low resource instrument for embarkation on any platform, including commercial satellites.

## 2. FIRST AMBER MEASUREMENTS

The first version of the Active Monitor Box of Electrostatic Risks (AMBER), designed as a double-head thermal electron and ion instrument, was launched onboard the Jason-3 (oceanography mission) spacecraft in 2016 [4]. This instrument was designed to measure ions, electrons and electrostatic charging on-board the spacecraft in its low-Earth orbit.

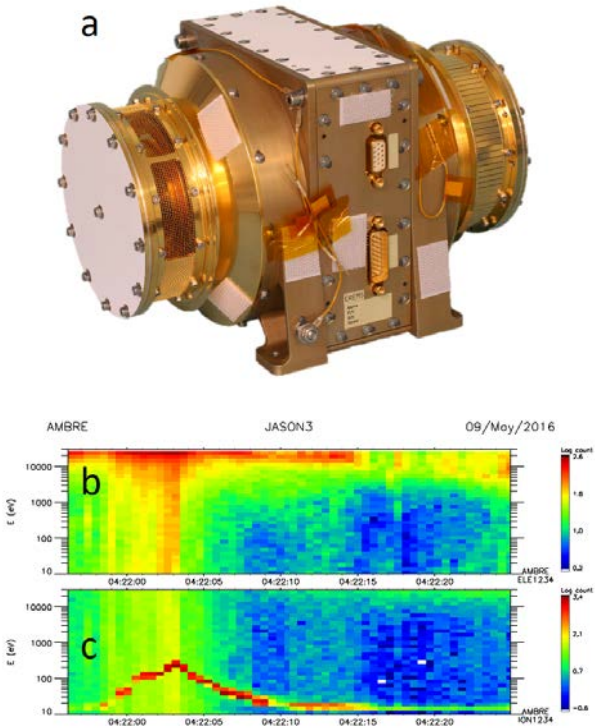

*Figure 1. (a) Picture of the AMBER instrument that is in flight on-board the Jason-3 spacecraft since 2016. (b) Electron and (c) ion energy-time spectrograms measured by AMBER on May 9, 2016. The high count rates (red color) measured at low energies in the ion spectrogram directly provide the information on the spacecraft electrostatic potential, which on this day went down to about -200 Volts..*

The electron and ion measurements rely on the use of two "top-hat" electrostatic analyser heads [5]. Positive and negative high voltage sweeps, respectively for electrons and ions, are applied on the inner hemisphere of the analysers, with a scanned energy range from few eV to 35 kilo-electron-Volts (keV). Each head comprises its own Micro-Channel Plates (MCP) chevron stack for particle detection at the exit of the electrostatic analyser. The electronics box, consisting of high voltage power supplies and an FPGA (Field-Programmable Gate Array) board to control the instrument, is located in between the two heads.

The instrument picture is shown in Fig. 1a, while Fig. 1b and 1c respectively show electron and ion measurements, in the form of energy – time spectrograms with color-coding displaying particle count rates (akin to energy fluxes). Figure 1c shows how AMBER provides the measurement of spacecraft electrostatic charging. Very large ion fluxes are observed at low energies as a thin (thus cold) red band (few eV to tens of eV). Such measurements are typical in the near-Earth magnetospheric region known as the plasmasphere, except that in principle these low energy ions should be confined to a few eV only. The reason why they are observed at higher energies during this pass, between 04:22:00 and 04:22:10 UT, is that they are accelerated towards the spacecraft surface by its own electrostatic potential to energies of several 100's of eV. Ion measurements in the form of such spectrogram therefore provide direct information on the spacecraft electrostatic charging, as the lower bound of these accelerated ions (cf. Fig. 1c).

The surface and electromagnetic cleanliness properties of the Jason-3 spacecraft typically lead to a negative spacecraft potential. On other platforms, however, depending on surface properties, orbit and environment (particle fluxes, eclipses, etc.) the surface of the spacecraft may charge positively. This is generally true for scientific spacecraft, which often require severe electromagnetic cleanliness constraints (conductive surfaces, etc.). In such cases, the spacecraft potential can be determined through electron measurements, with a break in the spectrum near the value of the spacecraft potential. This break in the spectrum stems from the very large fluxes of photoelectrons (from solar UV flux) below the spacecraft potential as they are trapped in the near-vicinity of the spacecraft by its potential (not shown; the reader can refer, e.g., to [6]). Being designed with both ion and electron measurement capabilities, the AMBER instruments (both the initial and new version) permit to measure both positive and negative spacecraft charging. It should be noted, however, that in the case of negative charging the electrostatic potential may be determined only in the presence of a low energy population (as observed in Fig. 1c.), such as the plasmasphere which is often present in the near-Earth magnetosphere.

## 3. NEW AMBER DEVELOPMENTS

The next generation AMBER instrument, for which a first prototype was developed and calibrated at the end of 2017 (Figures 2a and 2b), constitutes a significant evolution that is based on a single head to measure both particle species, ions and electrons, alternatively [7]. The instrument developments focused on several key sub-systems (mechanical design, front-end electronics, high-voltage electronics) that permit to reduce all instrument resources down to ~ 1 kg and 1.5 W.

First, a number of changes in the mechanical design were made to optimize the instrument overall resources (e.g., variable geometric factor discussed later). The mechanical design of the prototype instrument is shown in Fig. 2a.

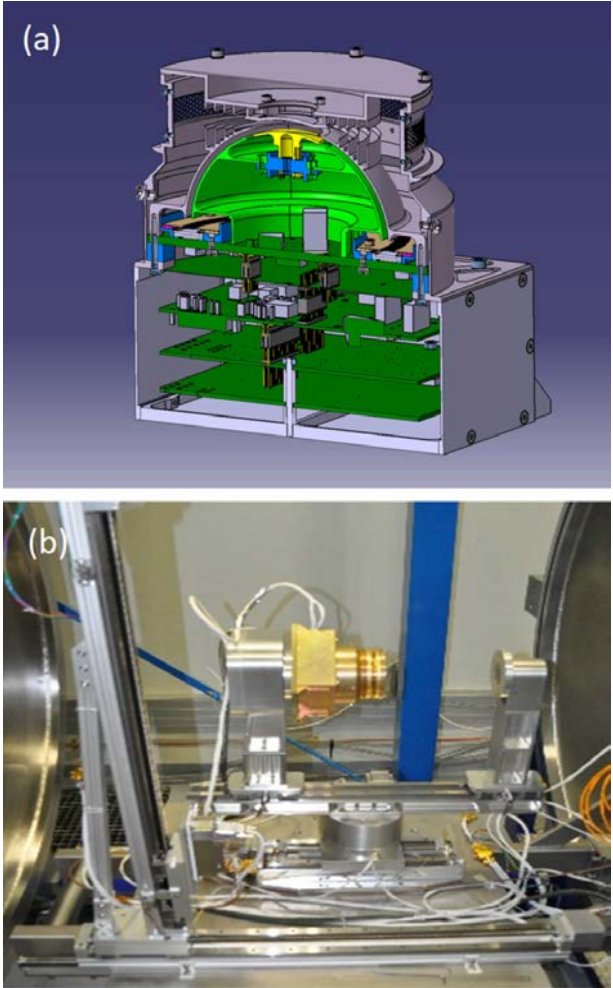

*Figure 2. (a) CAD mechanical drawing of the AMBRE 2 instrument prototype. (b) Picture of the AMBRE 2 prototype during calibration in a vacuum chamber at IRAP in 2017.*

An important part of the development then consisted in the development of new high voltage systems that permit both to alternate the electric potential applied on the inner hemisphere plate of the electrostatic analyser and, concomitantly, change the potential applied on the MCP detector stack, as required for each species. The details are not presented here, but a key improvement consisted in finding a high voltage system that does not strain the front-end electronics and MCP detectors too much. Indeed, the front-end design is sensitive to the changes in high voltages required to switch between species, and in particular to the rapidity of the switch, and is still being refined currently.

In order to further decrease resources, and at the same time increase the azimuthal resolution (from 45° on the first version of AMBER to 22.5° here) and the instantaneous azimuthal field-of-view (from 180° to 360°), we also developed a new compact front-end electronics board that uses a dedicated ASIC (Application Specific Integrated Circuit). This ASIC has been developed and space qualified by LPP for the Solar Orbiter mission. It includes 16 charged particles amplifiers with adjustable thresholds for noise discrimination and a test circuitry for on-board monitoring of the detection system. Our design benefits from its small footprint (1 cm²) and low power consumption (26 mW). This component has been also used for the electron instrument on-board the Parker Solar Probe mission [8].

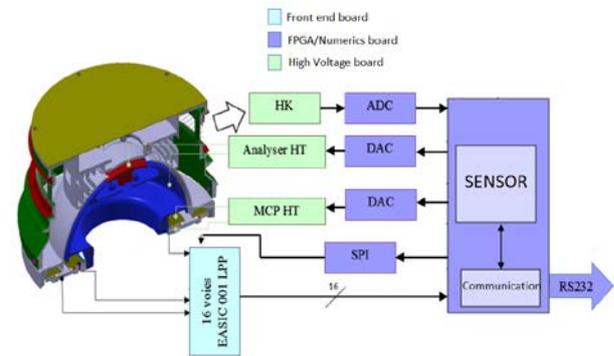

*Figure 3. Block diagram showing the instrument main sub-systems and their control during calibrations at IRAP. There is only one high voltage for the MCP stack but two high voltages for the analyser, one for the inner hemisphere and one for the top cap to vary the geometric factor.*

Table 1. Main properties of the new AMBER prototype.

| Parameter | Value |
| --- | --- |
| Sphere diameter | 86 mm |
| Azimuthal resolution (anode) | 22.5° |
| Azimuthal resolution (intrinsic) | 6° |
| Elevation resolution | 7.5° |
| Instantaneous FOV | 7.5° × 360° |
| Energy resolution | 14° |
| k-factor | 9.1 |
| Geometric factor (per anode) | 0.75 × 10$^{-3}$ cm².sr.eV/eV |
| Volume | 130 × 130 × 140 mm |
| Mass* | ~1 kg |
| Power* | ~1.5 W |

*\* Note that while the allocated volume would not change, the mass and power mentioned here do not account for the low voltage power supply (LVPS) board that is required to make AMBER stand-alone.*

In the Earth's magnetosphere, as in most other regions and planetary environments in the heliosphere, the thermal velocity of electrons is much higher than that of protons. This fact results, for particle analysers like that presented here, in much larger count rates for ions than electrons (about an order of magnitude in the Earth's magnetosphere). For that reason, while the analyser intrinsic geometric factor (cf. Table 1) is appropriate for electron count rates, ion fluxes would often lead to MCP detector saturation in a number of magnetospheric regions during intense conditions.

To mitigate this we implemented a variable geometric factor scheme, as previously designed in [9], by dividing the inner hemisphere into two anodes with different high voltage settings (Fig. 3.). This scheme allows us to decrease the geometric factor of the instrument by a factor up to 10 with a good level of precision. As an example of measurement that was made during the calibration of the prototype instrument, Fig. 4 shows the change in geometric factor as a function of the ratio of the voltages set on the upper and lower parts of the inner hemisphere plate. We show both the simulated and measured changes in geometric factor. A factor 0.6 exists between the two curves since the simulations do not account for overall detection efficiency (from grids, MCP, etc.).

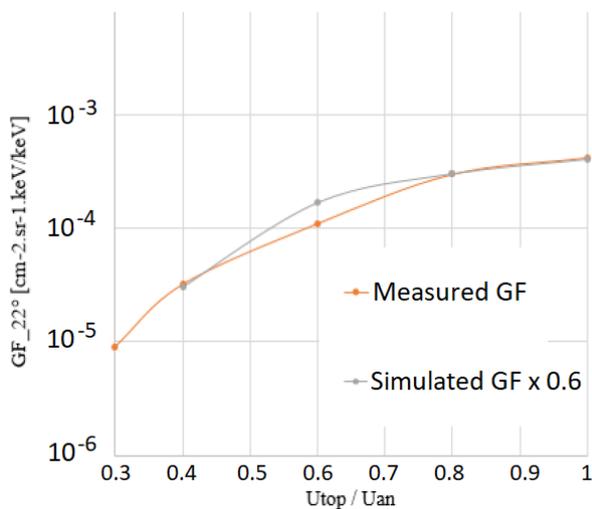

Figure 4. Instrument geometric factor variation as a function of the ratio between the voltages applied on the upper and lower parts of the inner hemisphere, from simulations (grey curve) and actual prototype calibrations in vacuum chamber with ion beam at IRAP.

### 4. FUTURE DEVELOPMENTS

Apart from the refinements that are ongoing regarding the front-end susceptibility to the rapidity of the switch between ion and electron measurements, most near-term efforts will focus on the design of flight-type FPGA, LVPS and communication electronics. Indeed, the FPGA board of the prototype instrument was made specifically for calibration purposes. Ongoing works thus focus on designing a more complete FPGA board, based on existing designs currently in flight (e.g., first version of AMBER). Ongoing and future works also focus on the design of the LVPS board, and the inclusion of versatile schemes for communication with the spacecraft bus, both in terms of TC/TM (telecommand/telemetry) and power connections.

Finally, although the current electrostatic analyser is based on a "top-hat" design [5] (Fig. 5a), a new electrostatic design called "cusp" [10] is being studied (Fig.5b). As presented in Fig. 5b, this design permits to improve the instrument, within the same overall resources (volume, mass and power), by increasing the energy range up to nearly 80 keV for ions.

### 5. CONCLUSIONS

AMBER is designed as a generic radiation monitor with a twofold purpose: (1) measure magnetospheric thermal ion and electron populations in the range 0-35 keV, with significant scientific potential (e.g., plasmasphere, ring current, plasma sheet), and (2) monitor spacecraft electrostatic charging and the plasma populations responsible for it, for electromagnetic cleanliness and operational purposes.

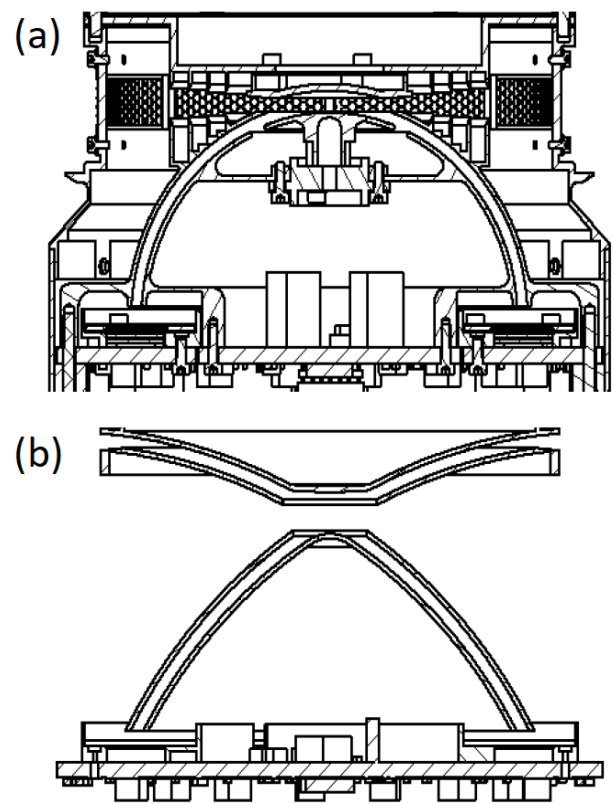

Figure 5. (a) Current design of the AMBER detection head with the classic "top-hat" analyser design. (b) "Cusp"-type analyser design that is envisaged for a future version of AMBER dedicated to the measurement of ions up to ~80 keV, within the same overall resources.

With its low resources (~1 kg and ~1.5 W), the current AMBER design, which was already prototyped and calibrated at the end of 2017, already constitutes a working version for embarkation on-board scientific or commercial spacecraft. Future developments are planned to adapt it for embarkation on any satellite.

## 6. ACKNOWLEDGEMENTS

This article is published in memory of Antoine Cara, who did the most part of the work all the way from the design to the building and calibration of the new version of AMBER [7]. Work at IRAP was supported by CNES, CNRS, and UPS. A. Cara's PhD work was supported by CNES and EREMS. The Aerospace Valley has in part supported recent developments and prospects.